\documentstyle[11pt,AATS,epsf,psfig]{article}	

\def\Mpc{\, h^{-1} \, {\rm Mpc}}
\def\bfw{ {\bf w}}

\def\Mpc{\ifmmode {\, h^{-1} \, {\rm Mpc}}
\else {$h^{-1}\,$ Mpc}\fi}

\def\s8{{\sigma_8}}

\def\ltsima{$\; \buildrel < \over \sim \;$}
\def\simlt{\lower.5ex\hbox{\ltsima}} 
\def\gtsima{$\; \buildrel > \over \sim \;$} 
\def\simgt{\lower.5ex\hbox{\gtsima}} 
\def\bfw{{\bf w}}

\def\omegam{{\Omega_{\rm m}}}
 
\def\omegab{{\Omega_{\rm b}}}
 
\def\omegal{{\Omega_\Lambda}}
\def\qrms{{Q_{\rm rms}}} 
\def\omegabh2{{\omegab h^2}} 

\begin{document}	

\title{The Age of the Universe from Joint Analysis of Cosmological Probes}

\author
{Ofer Lahav}
\affil{The Institute of Astronomy,
Madingley Road, Cambridge CB3 0HA,UK}


\begin{abstract}
Analyses of various cosmological probes, including the latest
Cosmic Microwave Background anisotropies, the 2dF Galaxy Redshift
Survey and Cepheid-calibrated distance indicators suggest that the age
of expansion is $13 \pm 3$ Gyr.  We discuss some statistical aspects
of this estimation, and we also present results for joint analysis of the
latest CMB and Cepheid data by utilizing `Hyper-Parameters'.  
The deduced age
of expansion might be uncomfortably close to the age of the oldest Globular
Clusters, in particular if they formed relatively recently.
\end{abstract}


\section{Introduction}

Estimating the age of the Universe in the framework of the Big Bang model 
is an old problem. The rapid progress in observational Cosmology 
in recent years has led to more accurate values 
of the fundamental cosmological parameters, including the age of the Universe. 
We summarize the basics in Section 2, we point out possible problems
in joint analysis of cosmological probes in section 3, and we summarize some 
recent results in Section 4.
In Section 5 we introduce a new method of `Hyper-Parameters' 
for combining different data sets, and we apply it in sections 6 and 7
to the latest Cepheid and CMB data. 
In section 8 we contrast the age of expansion
with the ages of Globular Clusters.






\section {The Age of Expansion}

In the standard Big Bang model 
the age of the Universe is found by integrating 
$dt = H^{-1} da/a$,  
where $a$ is the scale factor and 
$H \equiv {\dot a /a} $ is the Hubble parameter
as given by Einstein's equations.
This gives the present age in terms of three present-epoch parameters,
$H_0 = 100 h$ km/sec/Mpc = (9.78 Gyr$)^{-1}\;h$, 
the mass density parameter $\omegam$ and 
the scaled cosmological constant $\omegal
\equiv \Lambda/(3 H_0^2)$:
\begin{equation}
 t_0 = H_0^{-1} \; \int_0^1 \; da \; a^{1/2}\; 
[\omegal a^3 + (1-\omegam-\omegal)a +\omegam]^{-1/2} \;.
\end{equation}
The difficulty  is that in practice the  parameters ($H_0, \omegam, \omegal$),
when estimated from various cosmic probes, 
are commonly correlated with each other.
For a flat universe ($\omegam+\omegal =1$),
supported by the recent Cosmic Microwave Background (CMB) experiments, 
this integral has an analytic solution
in terms of only two free parameters:
\begin{equation}
t_0 = {2 \over 3} H_0^{-1} \; \omegal^{-1/2} \ln [(1 +\sqrt{\omegal}) (1-\omegal)^{-1/2}]
\end{equation}
which is well approximated 
(e.g. Peacock 1999) by:
\begin{equation}
 t_0 \approx  \; {2 \over 3} H_0^{-1} 
\omegam^{-0.3}\;. 
\end{equation}
This gives an insight to the way the errors propagate 
in the determination of the cosmic age:
\begin{equation}
{\Delta t_0 \over t_0 } \approx 
{\Delta H_0 \over H_0 } \; + \; 
0.3{\Delta \omegam \over \omegam }\;.
\end{equation}
This shows that the fractional error  in $H_0$ 
is about three times 
more important than the fractional error in $\omegam$.
Typically the quoted error on $H_0$ is 10\%
(e.g. Freedman et al 2000). The range of recent quoted
values for the density parameter suggests   
$\omegam \sim 0.3 \pm 50\% $, 
so the expected fractional error in age is about 25 \%
(e.g. for $t_0 \approx 13$ Gyr, $\Delta t_0 \approx 3$  Gyr).  
Again, $\omegam$ and $H_0$ are not always measured independently.
For example redshift survey constrains the shape 
of the Cold Dark Matter (CDM) 
power-spectrum via the product $\Gamma \equiv \omegam h$, 
while the CMB angular power-spectrum constrains
$\omega_{\rm m} = \omegam h^2$.

\section {Cosmological Parameters from a Joint Analysis: a Cosmic Harmony ? }

A simultaneous analysis of the constraints placed on  cosmological
parameters by different kinds of data is essential because
each probe (e.g. CMB,  
SNe Ia, redshift surveys, cluster abundance
and peculiar velocities)
typically constrains a different combination of
parameters. By performing joint likelihood analyses, one can
overcome intrinsic degeneracies inherent in any single analysis
and so estimate fundamental
parameters much more accurately. The comparison of
constraints can also provide a test for the validity of the assumed
cosmological model or, alternatively, a revised evaluation of the
systematic errors in one or all of the data sets.  Recent papers that
combine information from several data sets simultaneously include
Webster et al. (1998); Lineweaver (1998); 
Gawiser \& Silk (1998),  Bridle et al. (1999, 2001), 
Eisenstein, Hu \& Tegmark 1999; Efstathiou et al. 1999;  
and Bahcall et al. (1999). 

While joint Likelihood analyses employing both CMB and LSS data 
allow more accurate estimates of cosmological
parameters, they involve various subtle statistical issues:
\begin{itemize}

\item There is the uncertainty that a sample does not represent 
a typical patch of the FRW Universe to yield reliable global cosmological 
parameters.
\item The choice of the model parameter space is somewhat arbitrary.
\item One commonly solves for the probability for the data given a model
      (e.g. using a Likelihood function),  
      while in the Bayesian framework this should be modified
      by the prior for the model and its parameters.
\item If one is interested in a small set of parameters, should one marginalize
      over all the remaining parameters, rather than  fix them at certain 
      (somewhat ad-hoc) values ?  
\item The `topology' of the Likelihood contours may not be simple. 
      It is helpful when the Likelihood contours of different probes 
      `cross' each other to yield a global maximum 
       (e.g. in the case of CMB and SNe), but in other cases
       they may yield distinct separate `mountains', and the joint 
       maximum Likelihood may lie in a `valley'.
\item Different probes might be spatially correlated, i.e. 
       not necessarily independent.
\item What weight should one give to each data set ?
\end{itemize}

In a long term collaboration in Cambridge (Bridle
et al. 1999, 2001; Efstathiou et al. 1999; Lahav et al. 2000) we have
compared and combined in a self-consistent way the most powerful
cosmic probes: CMB, galaxy redshift surveys, galaxy cluster number
counts, type Ia Supernovae and galaxy peculiar velocities.  
Our analysis
suggests, in agreement with studies by other groups, that we live in a
flat accelerating Universe, with comparable amounts of dark matter and
`vacuum energy' (the cosmological constant $\Lambda$).

\section{Some Recent `Best Fit' Cosmological Parameters} 

To give the flavor
of favoured parameters we quote below two recent studies.
These and numerous other studies support a $\Lambda$-CDM model
with $\omegam = 1 - \omegal \sim 0.3$ and $h \sim 0.75$,
which corresponds to an expansion age of $t_0 \sim 12.6 $ Gyr
(and $H_0 t_0 =0.96$).

\subsection {Combining CMB, Supernovae Ia and Peculiar Velocities} 

A recent study (Bridle et al. 2001) is an example of 
combining 3 different data sets.
We compared and combined likelihood functions for the matter density
parameter $\omegam$, the Hubble constant 
$h$, 
and the normalization $\sigma_8$
(in terms of the variance in the mass density field measured in an
$8 h^{-1}$ Mpc radius sphere) 
from peculiar velocities, CMB (including the earlier Boomerang and Maxima data)
and type Ia
Supernovae. These three data sets directly probe the mass in the
Universe, without the need to relate the galaxy distribution to the
underlying mass via a ``biasing'' relation.  

Our analysis assumes a flat $\Lambda$-CDM cosmology with a
scale-invariant adiabatic initial power spectrum and baryonic fraction
as inferred from Big Bang Nucleosynthesis.  We find that all three
data sets agree well, overlapping significantly at the 2$\sigma$
level. This therefore justifies a joint analysis, in which we find a
best fit model and $95\%$ confidence limits of $\omegam =0.28\, (0.17,0.39)$, $h=0.74\, (0.64,0.86)$, and $\sigma_8=1.17\,
(0.98,1.37)$. In terms of the natural parameter combinations for these
data $\sigma_8\omegam^{0.6}=0.54\, (0.40,0.73)$, 
$\omegam h = 0.21\, (0.16,0.27)$. Also for the best fit point,
$Q_{\rm rms} = 19.7\mu$K and the age of the Universe is $13.0$
Gyr. 


\subsection {The 2dF Galaxy Redshift Survey}

The 2dF Galaxy Redshift Survey (2dFGRS) has now measured in excess of
160,000 galaxy redshifts and is the largest existing galaxy redshift
survey. A sample of this size allows large-scale structure statistics to
be measured with very small random errors. An initial analysis of
the power-spectrum
of the 2dFGRS (Percival et al. 2001) 
yields 68\% confidence limits on the total matter
density times the Hubble parameter $\omegam h = 0.20 \pm 0.03$, and the
baryon fraction $\Omega_b/\omegam = 0.15 \pm 0.07$, assuming
scale-invariant primordial fluctuations and a prior on the Hubble constant 
($h = 0.7 \pm 10 \% $).
Although the $\Lambda$-CDM model with comparable amounts of 
dark matter and dark energy is not so elegant, 
it is remarkable that various measurements show such good consistency.

\section{Hyper-Parameters}

We have  addressed recently (Lahav et al. 2000; Lahav 2001) the issue of
combining different data sets, which may suffer different systematic
and random errors.  We generalized the standard procedure of combining
likelihood functions by allowing freedom in the relative weights of
various probes.  This is done by including in the joint likelihood
function a set of `Hyper-Parameters', which are dealt with using
Bayesian considerations.  The resulting algorithm, which assumes
uniform priors on the logarithm of the Hyper-Parameters, is simple to
implement.
Here we show some examples of and results from the joint analysis of
the latest CMB and Cepheid data sets.

Assume that we have two independent data sets, $D_{A}$ and $D_{B}$
(with $N_{A}$ and $N_{B}$ data points respectively) 
and that we wish to determine a vector of free parameters ${\bfw}$
(such as the density parameter $\Omega_{\rm{m}}$, the Hubble constant $H_0$ etc.).
This is commonly done by minimizing  
\begin{equation}
\chi^2_{\rm joint} = \chi^2_A \; + \; \chi^2_B\; , 
\label{chi2_simple}
\end{equation}
(or, more generally,  maximizing 
the product of 
Likelihood functions).

Such procedures assume that the quoted observational random errors 
can be trusted, and that the two (or more) $\chi^2$s  
have equal weights.  
However, when combining `apples and oranges' one may wish to allow freedom in 
the relative weights. 
One possible approach is to generalize Eq. 5 to be 
\begin{equation}
\chi^2_{\rm joint} = \alpha  \chi^2_A \; + 
                \beta \; \chi^2_B \; , 
\label{chi2_hp}
\end{equation}
where $\alpha$ and $\beta$ are
 `Hyper-Parameters',
which are   to be dealt with 
the following Bayesian way.
There are a number of ways to interpret the meaning of the HPs.
One way is to understand $\alpha$ and $\beta$ as
controlling the relative weight of the two data sets.
It is not uncommon that astronomers 
accept and discard measurements (e.g. 
by assigning $\alpha=1$ and $\beta=0$)
in an ad-hoc way. 
The HPs procedure gives an objective diagnostic 
as to which measurements are problematic 
and deserve further understanding of systematic or random 
errors.   

How do we eliminate the unknown HPs  $\alpha$ and $\beta$ ?
This is done by marginalization over 
$\alpha$ and $\beta$ 
with Jeffreys' uniform priors in the log, 
$P(\ln \alpha) = P(\ln \beta) =1$.
We can then get
the probability for the parameters $\bfw$ given the data sets:
\begin{equation}
-2 \; \ln P(\bfw| D_{A}, D_{B})\; = 
N_{A} \ln (\chi_{A}^{2})   
\; + \;   N_{B} \ln (\chi_{B}^{2})\;. 
\label{lnPwDADB}
\end{equation} 
To find the  best fit parameters $\bfw$ requires us to minimize
the above probability in the $\bfw$ space.
It is as easy to calculate this statistic as the standard $\chi^2$, 
and it can be generalized for any number of data sets.

Since $\alpha$ and $\beta$
have been eliminated from the analysis by
marginalization they do not have particular values that can be quoted.
Rather, each value of $\alpha$ and $\beta$ has been considered and weighted
according to the probability of the data given the model.
It can be shown that the `weights' are
$ \alpha_{\rm {eff}} = \frac { N_{A}} {\chi_{A}^{2}}$ 
and
$\beta_{\rm {eff}} = \frac { N_{B}} {\chi_{B}^{2}} $,
both evaluated at the joint peak.

\section {$H_0$ from Cepheids}

One of the most important results for the Hubble constant comes
from the Hubble Space Telescope Key Project (Freedman et al. 2000).
The method is based on luminosity-period 
Cepheid calibration of several secondary distance
indicators measured over distances of 400 to 600 Mpc. 
Freedman et al. 2000 (see also in this volume) 
combined the different measurements
by several  statistical methods 
and derived as the `final result' 
$H_0  = 72 \pm (3)_r \pm (7)_s$ km/sec/Mpc 
(1-sigma random and systematic errors).

Given the importance of this work, we have attempted to combine the data
by a different method, using the HPs.
We used the raw data given by Freedman et al. (2000)
for Surface Brightness Fluctuations (SBF), Supernovae Ia (SNIa), 
Tully Fisher (TF) and Fundamental Plane (FP).
To the random errors given in the tables we added (in quadrature) 
the quoted systematic errors.

The results are shown on the right top and bottom 
panels in Figure 1. We see that the four methods give 
a range of values for $H_0$, with the most discrepant result
being the FP ($H_0 = 88$ km/sec/Mpc).
However, using the HPs, the  most probable result, $H_0 = 73$ km/sec/Mpc, 
agrees well with the result of Freedman et al.
We also see that in 
this case the standard joint $\chi^2$ and the HPs 
give a very similar answer.
The resulting HPs (`weights') for SBF, TF, SN and FP are
3.4, 2.7, 1.9 and  0.5, respectively.

\begin{figure}
\protect\centerline{
\psfig{figure=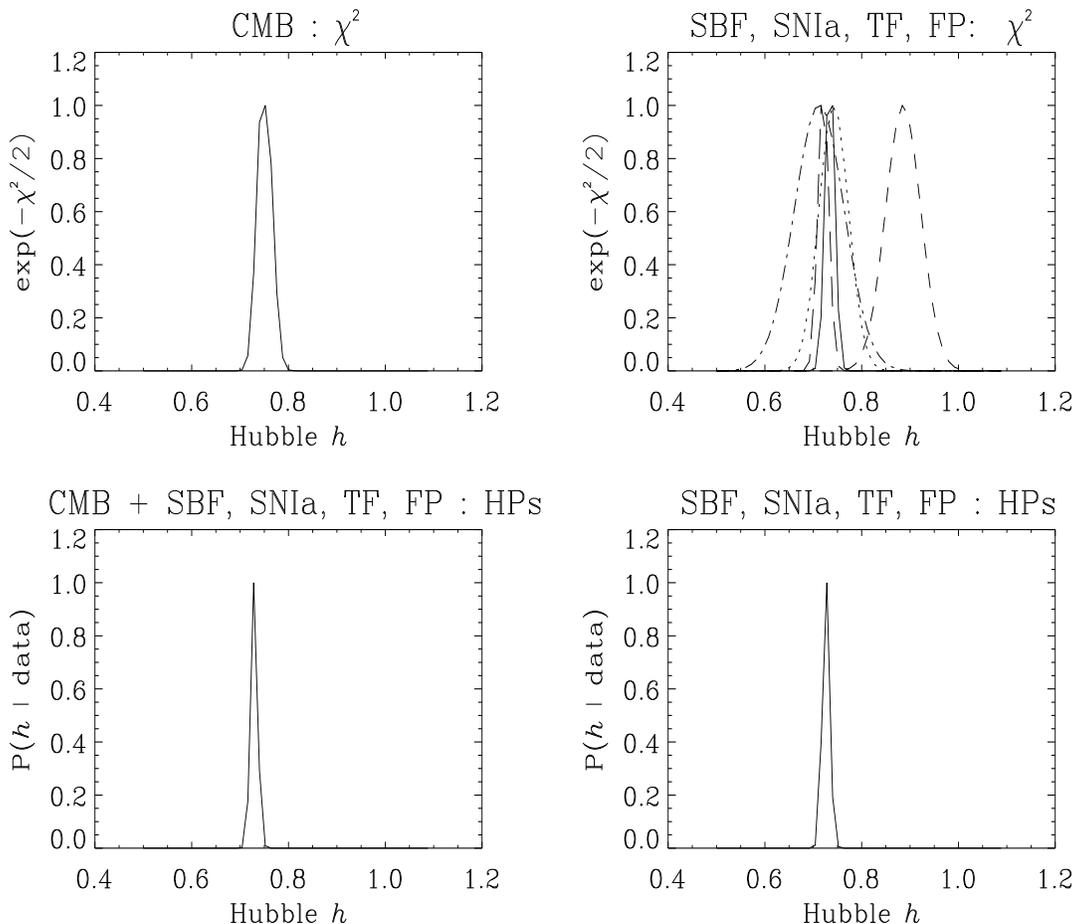,height=5.0truein,width=6.0 truein}}
\caption[]{
Probabilities for the Hubble constant.
{\bf Top right}: $\chi^2$ statistic for four Cepheid-calibrated
distance indicators from Freedman et al. (2000): SBF(dashed-dotted line) SNIa(long-dashed), 
TF(dotted) and FP(dashed). The joint $\chi^2$ for the four Cepheid data
is shown 
by the solid line, centred on $h=0.74$.
{\bf Bottom right}: The probability based on the Hyper-Parameters 
approach for the Hubble constant given the four Cepheid-calibrated data
set. The maximum probability is at $h=0.73$.
{\bf Top left}: 
$\chi^2$ statistic derived for  a compilation of the latest CMB data 
including 
Boomerang, Maxima and Dasi
from Wang et al. 2001) and a grid of CMB models 
for an assumed $\Lambda$-CDM model with 
with $n=1,
\omegam = 1 -\lambda = 0.3, \omegab h^2 = 0.02$ (BBN value), and 
$\qrms = 18 \mu K$ (COBE-normalization).
The maximum probability is at $h=0.75$ (see also Figure 2).
The distribution will be wider if some of the above parameters are kept free.
{\bf Bottom left }: 
The probability based on the Hyper-Parameters 
for the four Cepheid-calibrated data sets and the CMB compilation.
The maximum probability is at $h=0.73$.}
\end{figure}

\section {Combining Cepheids and  CMB  Data}


The latest Boomerang (Netterfield et al. 2001, de Bernardis et al. 2001),  
Maxima (Stomper et al. 2001) and Dasi (Pryke et al. 2001) CMB anisotropy measurements 
indicate 3 acoustic peaks.
Parameter fitting to a $\Lambda$-CDM model 
suggests consistency between the different experiments,
and a `best fit Universe' with zero curvature, 
and an initial spectrum with spectral index
$n =1$ (e.g. Wang et al. 2001 and references therein).
Unlike the earlier Boomerang and Maxima results, 
the new data also show that the  
baryon
contribution is consistent with the Big-Bang Nucleosynthesis
value $\omegab h^2 \sim 0.02$ (O'Meara et al. 2001).

\begin{figure}
\protect\centerline{
\psfig{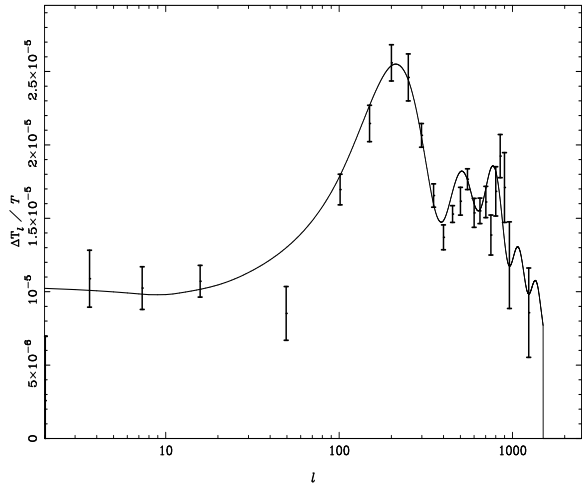}}
\caption[]
{A compilation of the latest CMB 
${\Delta T \over T}$ data points against spherical harmonic $l$
(from Wang et al. 2001). 
The line shows the predicted 
angular power-spectrum for a $\Lambda$-CDM model with $n=1,
\omegam = 1 -\lambda = 0.3, \omegab h^2 = 0.02$ (BBN value), 
$\qrms = 18 \mu K$ (COBE normalization)
and $h = 0.75$. A similar model (with $h=0.7$) is also 
the best fit to the 
the 2dF galaxy power-spectrum (Percival et al. 2001). 
Hence we see good agreement of 
two entirely different data sets.} 
\end{figure}

Several authors (e.g. Wang et al. 2001)
have pointed out that the Hubble constant $h$ itself
cannot be determined accurately from CMB data alone.
As the CMB constrains well the combination $\omega_{\rm m} \equiv \omegam h^2$,
the curvature $\Omega_k$ and $\omegal$, 
the Hubble constant can be derived from 
\begin{equation}
h = \sqrt{\omega_{\rm m} /(1-\Omega_k - \omegal)}.
\end{equation}
It is not surprising therefore that estimates of $h$ 
from the latest CMB data strongly depend on the assumed set of free
parameters, and on the assumed `priors' from other probes.
For example,  from the latest Boomerang data,
Netterfield et al. (2001) derive for `weak priors' $h=0.56 \pm 0.11$ 
and age $t_0 = 15.4 \pm 2.1 $ Gyr, while with `strong priors'
 ($h=0.66 \pm 0.05$; $t_0 = 14.0 \pm 0.6 $ Gyr).
Wang et al. (2001) find e.g. 
 ($h=0.42 \pm 0.23$;  $t_0 = 20.5 \pm 9.0 $ Gyr)
from CMB alone, 
and  ($h=0.57 \pm 0.30$ ; $t_0 = 14.2 \pm 4.3 $ Gyr)
by combining the CMB with the IRAS PSCz data.

Here, for simplicity, we take the approach of fixing all the other parameters,
apart from $h$.  
We assume that CMB fluctuations arise from
adiabatic initial conditions with Cold Dark Matter  and negligible
tensor component, in a flat Universe with  $\omegam=0.3$,
$\omegal  = 1-\omegam=0.7$, $n=1$,
$\qrms=18\mu$K (COBE-normalization) and $\omegab h^2 = 0.02$.
This choice is motivated by numerous other studies which combined
CMB data with other cosmological probes (e.g. 
Bridle et al. 2000; Hu et al. 2000; Wang et al. 2001; section 4 above).  
Of course, one may keep more free parameters, and marginalize
over some of them, as done in  numerous other studies.
We obtain theoretical CMB power-spectra using the CMBFAST and 
CAMB codes (Slejak \& Zaldarriaga 1996; Lewis, Challinor \& Lasenby 2000).
Increasing
$h$ decreases the height of the first acoustic peak, and makes few
other significant changes to the angular power spectrum (e.g. Hu et al. 2000). 
The range in $h$ investigated here is ($0.5<h<1.1$).

Different CMB data sets can be combined in different ways
(e.g. Jaffe et al. 2000; Lahav et al. 2000; Lahav 2001).
For simplicity we use here a compilation 
of 24 $\Delta T/T $ data points from Wang et al. (2001), 
which is based on 105 band-power measurements
(including the latest Boomerang, Maxima and Dasi).
The left top panel of Figure 1 shows the CMB likelihood 
function (with the correlation matrix not yet taken into account)
when the only free parameter is the Hubble constant.
It favours a value of $h \sim 0.75$.
Figure 2 shows the CMB data points, and we also projected 
our `best-fit' model (for the above set of assumptions).
A similar model (with $h=0.7$) also fits well 
other cosmological measurements, e.g. the 2dF galaxy power-spectrum
(Percival et al. 2001).  
The left bottom panel of Figure 1 shows the Hyper-parameters
joint probability for the CMB and the four Cepheid-calibrated data sets.
The maximum probability is at $h \sim 0.73$. 
We note that if more cosmological parameters are left free and then 
marginalized over, the error in $h$ would typically be much larger.
We also 
note that in the context of CDM models we can estimate the Hubble constant 
from the ratio 
\begin{equation}
h  = \omega_{\rm m}/\Gamma 
\end{equation}
(see section 2).
The recent CMB data suggest $\omega_{\rm m} \sim 0.15$ 
(e.g. Netterfield et al. 2001)
and from various 
redshift surveys $\Gamma \sim 0.2$,
so $h \sim 0.75$, in good agreement with our derived value. 
Of course $h$ and other parameters can be derived more quantitatively by 
joint likelihood 
analysis of CMB and redshift surveys (e.g. Webster et al. 1998).

\section{Discussion}

Joint analyses of cosmic probes suggests a flat Universe with
$\omegam=1-\omegal \approx 0.3 $.  The measurement of the Hubble
constant from Cepheids and from the CMB suggests $H_0 \approx 75 $
km/sec/Mpc.  This set of `best-fit' parameters yields an expansion age
$t_0 = 13 \pm 3$ Gyr.  
The error estimate  is due to errors in both $H_0$ and $\omegam$
(based on the range of recently quoted values). 
While this is currently the most popular model
there are potential problems with this set of parameters:

(i) There is no simple theoretical explanation why the present epoch 
contributions to matter $\omegam$ and 'dark energy' ($\omegal$) 
are nearly equal.

(ii) The age of expansion is commonly compared with the age
$t_{GC}$ of Globular Clusters (GC) and other old objects.
More precisely, one requires
\begin{equation}
t_{0} = t_{f} + t_{GC} \;, 
\end{equation}
where $t_{f}$ is the epoch of formation.

The age of the oldest 
GC was estimated to be $t_{GC} =11.5 \pm 1.3$ Gyr by Chaboyer
(1998), but he revised it upwards to $t_{GC} =13.2 \pm 1.5$ Gyr
(see this volume).
A recent radioactive dating (using $~^{238}$U, with half-life
of 4.5 Gyr) of a very metal-poor star in the Galaxy,
gives an age of $12.5 \pm 3 $ Gyr (Cayrel et al. 2001).

We see that the ages of old objects might be  uncomfortably close to the age
of expansion (despite having a non-vanishing cosmological constant,
which tends to stretch the age of the Universe).
Furthermore, it is commonly assumed that $t_f \sim 0.5 - 2$ Gyr
(e.g. Chaboyer 1998), and hence  $t_f$ is neglected in eq. (9).
However, we point out that $t_{f}$ is model dependent, and it can span
a wide range of values.  For example Peebles \& Dicke (1968) suggested
(before dark matter was recognized as a major component) that GC can
be identified with the Jeans' mass after re-combination, i.e. $z_f \sim
100-1000$ and indeed in this case $t_{f}$ is negligible.  

On the other hand, the model of Fall \& Rees (1985) for the formation of
GC in the Galactic halo as a result of thermal instability suggests
$z_f \sim 1-3$, i.e. $t_{f} \sim 2-5 $ Gyr for the above world model.
These and other models will
be discussed elsewhere (Gnedin, Lahav \& Rees, in preparation).
Obviously, late formation of the oldest GC means that the derived age of
expansion $t_0$ is too short compared with $t_{f} + t_{GC}$.
This possible `age crisis' 
might indicate a potential problem for either the cosmological model
or the age estimation of GC. 

The age of the Universe and other cosmological parameters will 
be revisited soon with larger and more accurate data sets
such as the big redshift surveys (2dF, SDSS) and CMB (MAP, Planck)
data.  
Other relevant probes for the Hubble constant are
the Sunyaev-Zeldovich effect (e.g. Mason et al. 2001), 
the gravitational lensing time delay (e.g. Williams \& Saha 2000)
and the baryon fraction in clusters (e.g. Ettori et al. 2001; Douspis et al.
2001).    
New high-quality data sets will allow us to study a wider
range of models and parameters.

\section*{Acknowledgments}

I thank Sarah Bridle, Pirin Erdogdu, Carolina \"Odman 
and the 2dFGRS team for their
contribution to the work presented here, and Wendy Freedman, Oleg
Gnedin, Jeremy Mould and Martin Rees for helpful discussions.  I also thank
Ted von Hippel and the other conference organizers for the hospitality in
Hawaii.

\end{document}